\documentclass[aps,pra,twocolumn,showpacs,floatfix]{revtex4}
\usepackage{graphicx}
\usepackage{dcolumn}
\usepackage{amsmath}
\usepackage{bm}

\begin{document}

\title{Finite-field evaluation of the Lennard-Jones atom-wall interaction
constant $C_3$ for alkali-metal atoms}

\author{W. R. Johnson}
\email{johnson@nd.edu} \homepage{www.nd.edu/~johnson}
\author{V. A. Dzuba}
\altaffiliation[Permanent address: ]{School of Physics, University
of New South Wales, Sydney 2052,
Australia}\email{V.Dzuba@unsw.edu.au}
\author{ U. I. Safronova}
\email{usafrono@nd.edu}
\affiliation{Department of Physics, 225 Nieuwland Science Hall\\
University of Notre Dame, Notre Dame, IN 46566}
\author{M. S. Safronova}
\affiliation{~223 Sharp Lab, Department of Physics and Astronomy\\
University of Deleware, Newark, DE 19716}

\date{\today}

\begin{abstract}
A finite-field scaling method is applied to evaluate the
Lennard-Jones interaction constant $C_3$ for alkali-metal atoms.
The calculations are based on the relativistic single-double (SD)
approximation in which single and double excitations of
Dirac-Hartree-Fock wave functions are included to all orders in
perturbation theory.
\end{abstract}


\maketitle

\section{Introduction}
The long-range interaction between an atom and a perfectly conducting wall
is governed by the \citet{LJ:32} static image potential
\begin{equation}
 V(z) = -\frac{e^2 C_3}{z^3} \, , \label{eq1}
\end{equation}
where $z$ is the distance between the atom and the wall.
The coefficient $C_3$ in Eq.~(\ref{eq1}) is the expectation value
of the operator
\[
 \frac{1}{16} \sum_{i,j} \left( x_i x_j + y_i y_j + 2 z_i z_j \right) \, ,
\]
in the atomic ground state.
Here, $\bm{r}_i = (x_i,\, y_i,\, z_i)$ is the coordinate of the
$i$-th atomic electron with respect to the nucleus. For an atom
with a spherically symmetric ground state, one can replace $C_3$
by the equivalent expression
\[
 C_3 = \frac{1}{12} \bigl< 0 \bigl| R^2 \bigr| 0 \bigr> \, ,
\]
where $\bm{R} = \sum_i \bm{r}_i $. The Lennard-Jones interaction
constant is important in models accounting for the finite
conductivity of the wall material by \citet{Bar:40} and
\citet{Mavr:63}. Additionally, the wall-atom-wall interaction
constant for small wall separation distances is proportional to
$C_3$ \cite{yan:97,karc:97}.

Precise values of $C_3$ for lithium
were obtained by \citet{yd:95} from an elaborate
configuration interaction (CI) calculation and confirmed by
an independent calculation by \citet{yan:97}.
The CI value of $C_3$ for lithium is in close agreement with the
value inferred from a variational calculation by King \cite{king:89}.
These accurate values of $C_3$ for lithium are about 2\% smaller
than the value obtained from a Hartree-Fock (HF) calculation.

An accurate semi-empirical value of $C_3$ for sodium was also obtained
by  \citet{karc:97} from an analysis of the $S_{-1}$ sum rule:
\[
  S_{-1} = \frac{2}{3} \langle R^2\rangle = \sum_{n} \frac{f_{n}}{E_0 - E_{n}},
\]
where $f_{n}$ is the oscillator strength of the transition from
the ground state to an excited state $n$. The quantities $E_0$ and
$E_{n}$ are energies of the ground state and excited state, respectively.
This value differs from the HF value of $C_3$ by about 10\%.
The more elaborate calculations by \citet{cff:96} improve the 
agreement between
theoretical and semi-empirical values for sodium somewhat.

Third-order many-body perturbation theory calculations of $C_3$ for
all alkali-metal atoms
and all-order singles-doubles calculations
of $C_3$ for Li, Na, and K were given by \citet{DJ:98}.
The all-order calculations for Li and Na were in close agreement with
with other precise values.  More recently, \citet{DJSB:99} deduced
accurate theoretical and semi-empirical values of $C_3$
for all alkali-metal atoms from oscillator-strength sum rules.

In the present work, we use finite-field many-body methods
to obtain values of $\left< R^2\right>$ and make comparisons with
previous work. One advantage of the finite field-method is that
all-order random-phase approximation (RPA) corrections 
are included from the start. A second
advantage is that n-th order corrections to the energy
give many-body perturbation theory (MBPT)
 corrections to matrix elements of $R^2$ normally associated 
with order (n+1).
A third advantage is that matrix elements of
one- and two-particle operators are essentially trivial to obtain,
in contrast to the lengthy calculations ordinarily required.

\section{Method}

The method used here to evaluate the expectation value of the
operator
\begin{equation}
R^2 = \biggl[ \sum_{i=1}^N \bm{r}_i \biggr]^2 = \sum_i r_i^2 + 2
\sum_{i\neq j} \bm{r}_i {\cdot}\, \bm{r}_j
\end{equation}
is known as the ``finite-field'' method and widely used in quantum
chemistry. We evaluate the ground-state expectation value of the
operator by adding the scalar operator $\lambda R^2$ to the
many-electron Hamiltonian and calculating the resulting energy
$E(\lambda)$. The desired expectation value is then given by
\[
\left< R^2\right> = \lim_{\lambda \to 0} \frac{dE}{d\lambda} .
\]
To evaluate the energy, we use standard many-body methods.

The modified many-electron Hamiltonian may be written
\[
H = \sum_i h_0(i)  + \frac{1}{2} \sum_{i\neq j} \left[
\frac{1}{r_{ij}} + 2 \lambda\, \bm{r}_i \bm{\cdot}\, \bm{r}_j
\right] - \sum_{i}  U(r_i) ,
\]
where,
\[
h_0 = c\, \bm{ \alpha \cdot p} + \beta mc^2 + V_\text{nuc}(r) +
\lambda\, r^2 + U(r).
\]
Expressing the Hamiltonian in second-quantized form and normally
ordering with respect to the closed core, we find
\begin{align}
H = E_0 +& \sum_{i} \epsilon_i :\! a^\dagger_i a_i\!\! : +
\frac{1}{2} \sum_{ijkl} v_{ijkl}
:\! a^\dagger_i a^\dagger_j a_l a_k \!\! : \nonumber \\
+& \sum_{ij} \left( V_\text{HF} -U\right)_{ij}:\! a^\dagger_i a_j
\!\! : ,
\end{align}
where
\[
E_0 = \sum_{a} \epsilon_a +  \sum_a \left( \frac{1}{2} V_\text{HF}
- U \right)_{aa} .
\]
In the above, equations,
\[
 \left( V_\text{HF} \right)_{ij} = \sum_a \left[ v_{iaja} - v_{iaaj} \right],
 \]
 with
 \[
 v_{ijkl} = \left< ij \left| \frac{1}{r_{12}}  + 2 \lambda {\bf r}_i {\bf \cdot}\, {\bf r}_j \right| kl \right> .
 \]

The lowest approximation to the wave function for a closed-shell
atom is
\[
  \Psi_0 = a^\dagger_a a^\dagger_b \cdots a^\dagger_n |0\rangle.
\]
The expectation of the Hamiltonian in this state, which gives the
first approximation to the ground-state energy, is
\[
\langle \Psi_0 | H | \Psi_0 \rangle = E_0 = \sum_{a} \epsilon_a +
\sum_a \left( \frac{1}{2} V_\text{HF} - U \right)_{aa} .
\]
In particular, if we choose $U = V_\text{HF}$, then the
corresponding value of the closed-shell energy is
\[
E_\text{HF} = \sum_{a}\left[  \epsilon_a - \frac{1}{2} \left(
V_\text{HF} \right)_{aa} \right] .
\]

\section{Results}
\subsection{HF-level} We consider an atom with one electron beyond closed
shells. The core and valence energies are given at the HF level of
approximation by
\begin{eqnarray}
E^\text{(HF)}_c &=& \sum_{a}\left[  \epsilon_a -
\frac{1}{2} \left(  V_\text{HF} \right)_{aa} \right] \\
E^\text{(HF)}_v  &=& \epsilon_v ,
\end{eqnarray}
where the summation index $a$ ranges over closed shells, and where
$\epsilon_v$ is the eigenvalue of the valence ``frozen core''
(Dirac)-Hartree-Fock equation.

In setting up the HF equations, we add $\lambda\, r^2$ to the
nuclear potential and add $2 \lambda\, {\bf r}_i {\bf \cdot}\,
{\bf r}_j$ to the two-particle interaction that defines the HF
potential. The modified electron-electron interaction becomes
\[
  \frac{1}{r_{12}} + 2 \lambda\, \bm{r}_1 \bm{\cdot}\, \bm{r}_2
  = \sum_{L=0}^\infty \frac{r_{\scriptscriptstyle <}^L}{r_{\scriptscriptstyle >}^{L+1}} P_L(\cos{\theta})
  + 2 \lambda\, r_1r_2\, P_1(\cos{\theta}).
\]
It follows that only the $L=1$ term in the HF potential is
modified and this term becomes
\begin{multline*}
v_1(ab,r) \rightarrow v_1(ab,r) \\ + 2 \lambda\, r \int_0^\infty\!\!
r' \left[ P_a(r')P_b(r') +  Q_a(r')Q_b(r')\right] dr' ,
\end{multline*}
where $P_a(r)$ and $Q_a(r)$ are large and small component radial
Dirac wave functions, respectively.
As a practical matter, we choose $\lambda \ll 1$
for neutral atoms to maintain some resemblance to the usual HF
picture. The HF energy includes first-order MBPT corrections,
together with all second- and higher-order RPA corrections. In
columns 2 and 3 of Table~\ref{tab1}, we list HF valence energies
$E_v^{\rm (HF)}(\lambda)$ and the HF core energies $E_c^{\rm
(HF)}(\lambda)$ as functions of $\lambda$ for the alkali-metal atoms
from Li to Fr.

\subsection{2nd-order MBPT}  We can easily go beyond the HF approximation
and include the second-order MBPT corrections to the core energy
$E^{(2)}_c$ and to the valence energy $E^{(2)}_v$
\begin{eqnarray}
E^{(2)}_c &=& -\frac{1}{2} \sum_{abmn} \frac{v_{mnab}\tilde{v}_{abmn}}
{\epsilon_m+\epsilon_n-\epsilon_a-\epsilon_b} \\
E^{(2)}_v &=& - \sum_{bmn} \frac{v_{mnvb}\tilde{v}_{vbmn}}
{\epsilon_m+\epsilon_n-\epsilon_v-\epsilon_b} \nonumber \\
&+& \sum_{abm} \frac{v_{mvab}\tilde{v}_{abmv}}
{\epsilon_m+\epsilon_v-\epsilon_a-\epsilon_b} .
\end{eqnarray}
In the above equations, indices $a$ and $b$ refer to core orbitals,
indices $m$ and $n$ refer to virtual orbitals, and $v$ refers to the
valence orbital.
The second-order energies
include those corrections to the matrix element of $R^2$
usually associated with third-order MBPT -- one interaction with $R^2$
and two Coulomb interactions.
In columns 4 and 5 of Table~\ref{tab1}, we list
$E_v^{\rm (2)}(\lambda)$ and $E_c^{\rm (2)}(\lambda)$ for various
values of $\lambda$.

\begin{table}
\caption{Variation of MBPT contributions to energies of
alkali-metal atoms with the scaling parameter $\lambda$, where
$H(\lambda) = H + \lambda R^2$.\label{tab1}}
\begin{ruledtabular}
\begin{tabular}{rrrrr}
\multicolumn{1}{c}{$\lambda$}& \multicolumn{1}{c}{$E_v^{\rm
(HF)}(\lambda)$}& \multicolumn{1}{c}{$E_c^{\rm (HF)}(\lambda)$}&
\multicolumn{1}{c}{$E_v^{(2)}(\lambda)$}&
\multicolumn{1}{c}{$E_c^{(2)}(\lambda)$}\\
\hline
\multicolumn{5}{c}{Li} \\
-0.00006& -0.1973918&    -7.2372588& -0.0016257&   -0.0400988 \\
-0.00004& -0.1970331&    -7.2372410& -0.0016327&   -0.0400995 \\
-0.00002& -0.1966760&    -7.2372232& -0.0016397&   -0.0401001 \\
 0.00002& -0.1959663&    -7.2371876& -0.0016535&   -0.0401014 \\
 0.00004& -0.1956137&    -7.2371698& -0.0016603&   -0.0401020 \\
 0.00006& -0.1952625&    -7.2371520& -0.0016671&   -0.0401026 \\
  \multicolumn{5}{c}{Na} \\
-0.00006& -0.1832811&  -161.8961446& -0.0057754&   -0.3836680 \\
-0.00004& -0.1828627&  -161.8960647& -0.0058073&   -0.3836734 \\
-0.00002& -0.1824466&  -161.8959849& -0.0058389&   -0.3836788 \\
 0.00002& -0.1816209&  -161.8958252& -0.0059014&   -0.3836896 \\
 0.00004& -0.1812111&  -161.8957454& -0.0059323&   -0.3836950 \\
 0.00006& -0.1808033&  -161.8956655& -0.0059630&   -0.3837004 \\
       \multicolumn{5}{c}{K} \\
-0.00006& -0.1493919&  -601.3789128& -0.0120214&   -0.7237274 \\
-0.00004& -0.1487522&  -601.3786754& -0.0121561&   -0.7237770 \\
-0.00002& -0.1481185&  -601.3784379& -0.0122889&   -0.7238267 \\
 0.00002& -0.1468683&  -601.3779630& -0.0125494&   -0.7239262 \\
 0.00004& -0.1462511&  -601.3777256& -0.0126774&   -0.7239760 \\
 0.00006& -0.1456389&  -601.3774882& -0.0128041&   -0.7240257 \\
  \multicolumn{5}{c}{Rb} \\
-0.00006& -0.1414238& -2979.6664077& -0.0144304&   -1.8931869 \\
-0.00004& -0.1407050& -2979.6660442& -0.0146265&   -1.8932719 \\
-0.00002& -0.1399943& -2979.6656808& -0.0148194&   -1.8933569 \\
 0.00002& -0.1385951& -2979.6649539& -0.0151967&   -1.8935271 \\
 0.00004& -0.1379058& -2979.6645905& -0.0153817&   -1.8936122 \\
 0.00006& -0.1372228& -2979.6642272& -0.0155647&   -1.8936974 \\
      \multicolumn{5}{c}{Cs} \\
-0.00006& -0.1299267& -7786.6477893& -0.0168211&   -3.1079015 \\
-0.00004& -0.1290615& -7786.6472240& -0.0171307&   -3.1080755 \\
-0.00002& -0.1282090& -7786.6466589& -0.0174334&   -3.1082495 \\
 0.00002& -0.1265379& -7786.6455289& -0.0180225&   -3.1085978 \\
 0.00004& -0.1257176& -7786.6449640& -0.0183103&   -3.1087722 \\
 0.00006& -0.1249066& -7786.6443992& -0.0185943&   -3.1089466 \\
  \multicolumn{5}{c}{Fr} \\
-0.00006& -0.1334806&-24307.9714413& -0.0206103&   -5.8273192\\
-0.00004& -0.1326676&-24307.9707289& -0.0209570&   -5.8275533\\
-0.00002& -0.1318658&-24307.9700168& -0.0212968&   -5.8277875\\
 0.00002& -0.1302922&-24307.9685926& -0.0219595&   -5.8282563\\
 0.00004& -0.1295191&-24307.9678807& -0.0222839&   -5.8284909\\
 0.00006& -0.1287542&-24307.9671690& -0.0226043&   -5.8287257\\
\end{tabular}
\end{ruledtabular}
\end{table}

\subsection{Differentiation Formulas}
The energies are given on the grid
\[
\lambda_n = (-3h,\, -2h,\,  -h,\, 0,\, h,\, 2h,\, 3h)
\]
with spacing $h = 0.00002$.
To evaluate $\left<R^2\right>$, we make use of an hierarchy of successively
more accurate Lagrangian differentiation formulas

\begin{align*}
\left( \frac{dE}{d\lambda }\right) ^{(3)} =&\ \frac{1}{2h}\Bigl(
E[1]+E[-1]\Bigr)  \\
\left( \frac{dE}{d\lambda }\right) ^{(5)} =&\ \frac{1}{24h}\Bigl(
16(E[1]-E[-1]) \Bigr.\\
& \qquad \Bigl. -2(E[2]-E[-2]) \Bigr)  \\
\left( \frac{dE}{d\lambda }\right) ^{(7)} =&\ \frac{1}{720h}\Bigl(
540(E[1]-E[-1])\Bigr.  \\ 
& \Bigl. -108(E[2]-E[-2])+12(E[3]-E[-3]) \Bigr)
\end{align*}
to obtain $dE/d\lambda$ at $\lambda=0$. In the above, we designate
$E(\lambda_n)$ by $E[n]$. The k-th approximation to the derivative 
$\left(dE/d\lambda\right)^{(k)}$ has an error proportional to
$h^k$.  The first two of these formulas are given 
in \citet[][Chap.\ 25]{AS:68}.

In Table~\ref{tab2}, we show results of applying the differentiation 
formulas to
the data in Table~\ref{tab1}. The resulting values of $\left<R^2\right>$
are numerically stable to about 4 digits for the cases considered.
These values are compared with values from third-order MBPT and other
accurate values in Table~\ref{tab3}.
\begin{table}
\caption{Values of $(dE/d\lambda)_{\lambda=0} =\left< R^2\right>$
for alkali-metal atoms
 as order of differentiation formula is increased. Step size: $h = 0.00002$.
\label{tab2}}
\begin{ruledtabular}
\begin{tabular}{crrrrr}
\multicolumn{1}{c}{Order} & \multicolumn{1}{c}{$dE_v^{\rm
(HF)}/d\lambda$}& \multicolumn{1}{c}{$dE_c^{\rm (HF)}/d\lambda$}&
\multicolumn{1}{c}{$dE_v^{(2)}/d\lambda$}&
\multicolumn{1}{c}{$dE_c^{(2)}/d\lambda$}&
\multicolumn{1}{c}{$\left< R^2\right>$}\\
\hline
       \multicolumn{6}{c}{Li}\\
3& 17.7418&  0.8904&  -0.3445& -0.0315& 18.256\\
5& 17.7415&  0.8904&  -0.3445& -0.0315& 18.256\\
7& 17.7415&  0.8904&  -0.3445& -0.0315& 18.256\\
  \multicolumn{6}{c}{Na}\\
3&  20.6433&  3.9922&  -1.5631& -0.2694& 22.803\\
5&  20.6427&  3.9922&  -1.5629& -0.2694& 22.803\\
7&  20.6427&  3.9922&  -1.5629& -0.2694& 22.803\\
  \multicolumn{6}{c}{K}\\
3&  31.2556& 11.8718&  -6.5118& -2.4864& 34.129\\
5&  31.2531& 11.8718&  -6.5105& -2.4864& 34.128\\
7&  31.2532& 11.8718&  -6.5105& -2.4864& 34.128\\
  \multicolumn{6}{c}{Rb}\\
3&  34.9792& 18.1725&  -9.4361& -4.2549& 39.461\\
5&  34.9755& 18.1729&  -9.4336& -4.2549& 39.460\\
7&  34.9755& 18.1730&  -9.4337& -4.2549& 39.460\\
  \multicolumn{6}{c}{Cs}\\
3&  41.7773& 28.2500& -14.7337& -8.7108& 46.583\\
5&  41.7703& 28.2499& -14.7276& -8.7108& 46.582\\
7&  41.7704& 28.2498& -14.7278& -8.7108& 46.582\\
  \multicolumn{6}{c}{Fr}\\
3&  39.3401& 35.6056& -16.5691&-11.7207& 46.656\\
5&  39.3343& 35.6068& -16.5634&-11.7207& 46.657\\
7&  39.3344& 35.6073& -16.5635&-11.7207& 46.658\\
\end{tabular}
\end{ruledtabular}
\end{table}

\begin{table}
\caption{Comparison of the present second-order finite-field
(FF$^{(2)}$) values of $R^2$ with third-order MBPT values from
\cite{DJ:98} and with semi-empirical (SE) values from
\cite{DJSB:99}. The value for Li reported under SE is obtained by
rounding the ``exact'' value 18.216$\ldots$ given in
Refs.~\cite{king:89,yd:95}. \label{tab3}}
\begin{ruledtabular}
\begin{tabular}{cccc}
 \multicolumn{1}{c}{Element} &
\multicolumn{1}{c}{FF$^{(2)}$} & \multicolumn{1}{c}{MBPT} &
\multicolumn{1}{c}{SE} \\
\hline
Li & 18.26 & 18.26 & 18.22\\
Na & 22.80 & 22.79 & 22.45\\
K~ & 34.13 & 34.05 & 34.52\\
Rb & 39.46 & 39.37 & 40.92\\
Cs & 46.59 & 46.35 & 50.96\\
\end{tabular}
\end{ruledtabular}
\end{table}

\subsection{Third-Order MBPT \label{subd}}

Expressions for third-order correlation corrections to
core and valence energies of atoms with a single valence
electron were given in \cite{BGJS:87} and applied to
study ground-state removal energies of Cs and Tl in \cite{BJS:90}.
In the present applications, these formulas are used to
evaluate, effectively, fourth-order corrections to matrix elements
of $R^2$. Although we do not expect the third-order calculations
presented in this section to be as accurate as the singles-doubles (SD)
calculations given in the
following section, it is in any case necessary to carry
out third-order energy calculations to determine  $E^{(3)}_\text{extra}$,
the correction to the SD energies that accounts approximately for
omitted triple excitations in the SD equations.

Third-order corrections for lithium are
$dE^{(3)}_v/d\lambda = -0.0297$ and $dE^{(3)}_c/d\lambda = -0.0011$.  
Adding these values to the earlier second-order result,
leads to $\left<R^2\right> = 18.2250$ for lithium. This slightly
improves the agreement of MBPT with the exact nonrelativistic value.
However, better agreement can be achieved in the SD approximation.
Therefore, here we only calculate $dE^\text{(3)}_\text{extra}/d\lambda$.
These contributions are 0.00252 for Li, 0.00201 for Na, 0.72532 for K,
1.10114 for Rb, 2.32450 for Cs, and 2.59698 for Fr.

\subsection{All-Order Singles-Doubles}
The singles-doubles (SD) equations, also referred to as all-order pair
equations \cite[][Chap. 15]{LM:86}, provide a method of including
important correlation corrections to the atomic wave function
to all-orders in perturbation theory.
One solves a set of coupled equations
for single excitation  coefficients
$\rho_{ma}$, $\rho_{mv}$ and double excitation coefficients $\rho_{mnab}$,
$\rho_{mnva}$ of the HF ground state \citep[see][for example]{SDJ:98}.
Once these expansion coefficients have been determined,
the correlation correction to the core energy $\Delta E_c$ is given by
\begin{equation}
\Delta E_c =
  \frac{1}{2} \sum_{mnab} v_{abmn}\tilde{\rho}_{mnab} . \label{enc}
\end{equation}
and the correlation correction to the valence energy is given by
\begin{eqnarray}
\Delta E_v &=& \sum_{ma} \tilde{v}_{vavm}\rho_{ma} \\
&+& \sum_{mab} v_{abvm}\tilde{\rho}_{mvab}
+ \sum_{mna} v_{vbmn}\tilde{\rho}_{mnvb} . \label{env} \nonumber
\end{eqnarray}
The core energy is exact through third order in MBPT and contains
important fourth- and higher-order corrections. The valence energy
also includes important fourth- and higher-order corrections but
is missing small third-order corrections 
\citep[written out explicitly in][]{SDJ:98} referred to as $E^{(3)}_\text{extra}$.
These missing terms have their origin in omitted triple excitations
(single-valence -- double-core excitations)
of the HF ground state.  Numerical values of $E^{(3)}_\text{extra}$
for the alkali-metal atoms are given at the end of Section \ref{subd}. 

\paragraph{Lithium:} Calculations for Li 
include all partial waves with $l \leq 6$. To estimate higher $l$
contributions we use Aitken's $\delta^2$ method.
Table~\ref{tab8} shows
contributions to $\left< R^2\right>$ evaluated
with $l_\text{max}$ ranging from 2 to 5. 
The final extrapolated value $\left<R^2\right>=18.213$ from Table~\ref{tab8}
differs from the ``exact'' nonrelativistic value (18.216004)
for lithium given by \citet{yd:95},
but is in precise agreement with an earlier
SD result by~\citet{DJ:98}.
The small difference with the exact nonrelativistic value is dominated
by the contribution from $E^{(3)}_\text{extra}$, evaluated in the
previous subsection, which has the value
$dE^\text{(3)}_\text{extra}/d\lambda = 0.0025$.
When this correction is added to the SD result 18.2130 for lithium,
we obtain the value 18.2155, differing  from the exact
nonrelativistic result by only -0.0005. The residual difference has the sign
and order of magnitude expected for a relativistic correction
to $R^2$.

\begin{table}
\caption{Calculated values of $\left<R^2\right>$ for Li 
 as the number of partial waves $l_\text{max}$ included in the SD equations
is increased are tabulated along with extrapolated values obtained
by applying Aitken's $\delta^2$ method to $l_\text{max} = (2,\,
3,\, 4)$ and $(3,\, 4,\, 5)$. \label{tab8}}
\begin{ruledtabular}
\begin{tabular}{crrr}
      \multicolumn{1}{c}{$l_\text{max}$} &
      \multicolumn{1}{c}{$d\Delta E_v/d\lambda$}&
      \multicolumn{1}{c}{$d\Delta E_c/d\lambda$}&
      \multicolumn{1}{c}{$\left<R^2\right>$}\\
      \hline
    2&  -0.37905&   -0.03284&    18.2200\\
    3&  -0.38362&   -0.03279&    18.2155 \\
    4&  -0.38509&   -0.03279&    18.2140\\
    5&  -0.38570&   -0.03279&    18.2134 \\
\hline
2-3-4&  -0.38579&   -0.03279&    18.2133 \\
3-4-5&  -0.38612&   -0.03279&    18.2130 \\
\end{tabular}
\end{ruledtabular}
\end{table}

\paragraph{Other alkalis}
In Table~\ref{tab9},  we show the derivatives of valence and core energies 
of alkali atoms from Li to Fr calculated in the SD approximation
with $l_\text{max}=6$  as the order of the differentiation is increased.
We also include the contribution from the missing third-order
energy $E^{(3)}_\text{extra}$ evaluated in the previous section.

The SD result for sodium  $\left< R^2\right> = 22.6425(3)$ agrees well
with the earlier SD result 22.6293 from \citet{DJ:98} and with the
semi-empirical value 22.65 from  \citet{karc:97}.
Note however, that present results for all alkali atoms other than lithium
are substantially larger than semi-empirical values obtained in
Ref.~\cite{DJSB:99} (see Table~\ref{tab3}).

The resulting values $\left<R^2\right>$ from the SD calculation,
which are our most accurate predictions, are summarized in Table~\ref{tab10}.

\begin{table}
\caption{Values of $(dE/d\lambda)_{\lambda=0}$ in the SD+$E^{(3)}_\text{extra}$
approximation
as the order of differentiation formula is increased. Step size: $h = 0.00002$.
The SD equations included all partial waves with $l\leq 6$ for Li, Na, K, Rb, Cs
 and with $l\leq 5$ for Fr.
\label{tab9}}
\begin{ruledtabular}
\begin{tabular}{crrrrrr}
\multicolumn{1}{c}{Order} &
\multicolumn{1}{c}{$\frac{dE_{v}^{\rm (HF)}}{d\lambda }$}&
\multicolumn{1}{c}{$\frac{dE_{c}^{\rm (HF)}}{d\lambda }$}& 
\multicolumn{1}{c}{$\frac{d\Delta E_{v}}{d\lambda }$}&
\multicolumn{1}{c}{$\frac{d\Delta E_{c}}{d\lambda }$}&
\multicolumn{1}{c}{$\frac{dE_{\rm extra}^{(3)}}{d\lambda }$ }&
\multicolumn{1}{c}{$\left<R^2\right>$}\\
\hline \multicolumn{7}{c}{Li}\\
3&  17.7418&  0.8904&   -0.3860&  -0.0328&  0.0025&  18.216 \\
5&  17.7415&  0.8904&   -0.3859&  -0.0328&  0.0025&  18.216 \\
7&  17.7415&  0.8904&   -0.3859&  -0.0328&  0.0025&  18.216\\
\multicolumn{7}{c}{Na}\\
3&  20.6433&  3.9922&   -1.7038&  -0.2922&  0.0020&  22.642 \\
5&  20.6427&  3.9922&   -1.7037&  -0.2922&  0.0020&  22.641 \\
7&  20.6427&  3.9922&   -1.7037&  -0.2922&  0.0020&  22.641\\
\multicolumn{7}{c}{K}\\
3&  31.2556&  11.8719&  -5.9711&  -2.2866&  0.7255&  35.595 \\
5&  31.2531&  11.8718&  -5.9703&  -2.2866&  0.7253&  35.593 \\
7&  31.2532&  11.8718&  -5.9703&  -2.2866&  0.7253&  35.593\\
\multicolumn{7}{c}{Rb}\\
3&  34.9792&  18.1725&  -8.2136&  -3.6888&  1.1014&  42.351 \\
5&  34.9755&  18.1729&  -8.2119&  -3.6888&  1.1011&  42.349 \\
7&  34.9755&  18.1730&  -8.2117&  -3.6888&  1.1011&  42.349\\
\multicolumn{7}{c}{Cs}\\
3&  41.7773&  28.2500& -11.6972&  -6.6590&  2.3254&  53.997 \\
5&  41.7703&  28.2499& -11.6942&  -6.6590&  2.3245&  53.991 \\
7&  41.7704&  28.2498& -11.6943&  -6.6590&  2.3245&  53.991 \\
\multicolumn{7}{c}{Fr}\\
3&  39.3401&  35.6056& -12.4105&  -8.5968&  2.5970&  56.536 \\

\end{tabular}
\end{ruledtabular}
\end{table}

\section*{Conclusion}

In this paper we present the most complete fully {\it ab initio} all-order
calculations of the Lennard-Jones interaction constant $C_3$ for alkali-metal
atoms. Incorporating of the rescaled $R^2$ operator into original Hartree-Fock
Hamiltonian allows us to stay within standard SD technique while also including
important subclasses of higher-order contributions.
Results for Li agree precisely with the ``exact'' CI results of \citet{yd:95}, 
while results for other alkali atoms are probably the most accurate available 
to date.

\begin{table}[t]
\caption{Final results for $\left<R^2\right>$ for alkali atoms.
\label{tab10}}
\begin{ruledtabular}
\begin{tabular}{cccccc}
 Li & Na & K & Rb & Cs & Fr \\
\hline
18.216& 22.641& 35.593& 42.349& 53.991& 56.536 \\
\end{tabular}
\end{ruledtabular}
\end{table}

\section*{Acknowledgment}

The work of W.R.J. and U.I.S was supported in part by 
National Science Foundation Grant
No.\ PHY-01-39928.


\end{document}